\documentclass{article}
\usepackage[usenames,dvipsnames]{pstricks}
\usepackage{epsfig}
\usepackage{pst-grad}
\usepackage{pst-plot}

\begin{document}
\title{\bf On the elements of the Earth's\\ ellipsoid of inertia}
\author{Alina-Daniela  V\^{I}LCU}
\date{}
 \maketitle

 \textbf{Abstract}- By using the data for the known geopotential
models by means of
 artificial satellite, the central moments of inertia of the Earth are determined. For this purpose, it was used
 the value $H = 0.00327369\pm9.8\cdot10^{-8}$ for dynamical
 flattening of the Earth \cite{MIH}. The results obtained indicate that the pole
 of inertia is located near the Conventional International Origin (CIO). Also, the orientation of the triaxial
 ellipsoid of inertia for nine geopotential models considered is given.
 Our results improve the ones obtained by Erzhanov and
 Kalybaev \cite{EK}.
\endabstract

\section{Introduction}

The first artificial satellite of the Earth, Sputnik-1, was launched
on 4 October 1957. Studying the trajectory of the following
satellites (Sputnik-2 and Sputnik-3), D. King-Hele has determined
the zonal coefficient $J_{2}$. The value $J_{2}=1.084\cdot10^{-3}$
was quite close to that calculated by terrestrial  measurements. The
US satellite, Vanguard-1, launched in March 1958, made it possible
for the first time the assessment of discrepancy between the
ellipsoid and geoid. The value of $J_{4}$ was obtained in the same
year and the first odd zonal term in 1959 by Y. Kozai. In 1961, W.
Kaula produced a complete model of degree 4, involving all the
coefficients $C_{lm}$ and $S_{lm}$ of the associated Lagrange
function $P_{22}$. From this moment, the information about the
gravitational field of the Earth has become more numerous and
accurate.

The first data from satellites have been used in the development of
geopotential models from the early 1970. The SAO-SE model
(Smithsonian Astrophysical Observatory - Standard Earth),
established in 1966, used in 1972 the first laser-ranging
measurements to establish satellite distances. The GEM model
(Goddard Earth Model) was established by NASA's GSFC (Goddard Space
Flight Center) in the United States as a reaction to the classified
US military models. The first model, GEM-1, was published in 1972,
expanding the potential to degree 12. He then followed the whole
series of geopotential models to the GEM-10 (developed up to order
20). Subsequently it was developed the model EGM, as a result of the
collaboration between GSFC-NASA, NIMA (National Imagery and Mapping
Agency) and OSU (Ohio State University). In 1996 came EGM96S, of
degree and order 70, with data provided solely by satellites, and
EGM96, of degree and order 360, adjoining geophysical data
\cite{CAP}.

In this paper, using the data provided by the SE-2 geopotential
models from SE series, GEM-5 to GEM-10 models from GEM series and
the EGM96 model, the Earth's moments of inertia are calculated.
Although at most six digits are accurate, to compare the results
more easily with the values obtained by Erzhanov and Kalybaev
\cite{EK}, the calculations are performed with nine digits. In order
to determine the polar moment $C'$ it was used the equation obtained
by Prof. Ieronim Mihaila \cite{MIHA}. In fact, for all the nine
models used, the value of $C'$ in Tables 4 and 6 coincides with the
value of $C'$ obtained by considering
\[H'=\frac{1}{2C'}[2C'-(A'+B')]=H\] like in \cite{EK}, where the dynamical
flattening of the Earth
\[H=\frac{1}{2C}[2C-(A+B)]\] is obtained from the constant of precession.
We used here for dynamical flattening the value $H =
0.00327369\pm9.8\cdot10^{-8}$ \cite{MIH}. Thus, it demonstrates that
the choice made in \cite{EK}, namely $H'=H$, is valid until the
order of $10^{-9}$. The orientation of the ellipsoid of inertia is
described in Table 7. The data from satellites on the gravitational
potential indicate that the equatorial principal moments of inertia
of the Earth are not equal (see Table 6) and it also shows us that
the polar axis does not coincide with the axis of rotation. The pole
of inertia $P_{i}$ remains near the Conventional International
Origin.

\section{Representations of the geopotential }

In the theory of the movement of the Earth's artificial satellites
it is chosen as a reference system the geocentric system
$O\xi\eta\zeta$, the axis $O\zeta$ of the system being given by the
position of the Conventional International Origin. The origin plan
for longitude, $O\xi\zeta$, is the plan of the Greenwich meridian.

In polar coordinates, the expression of the geopotential is
\cite{EK}
\begin{eqnarray}
U(r,\varphi,\lambda)&=&G\sum_{n=0}^{\infty}\frac{1}{r^{n+1}}\sum_{m=0}^{n}\frac{2}{\delta_{m}}\frac{(n-m)!}{(n+m)!}
P_{n}^{(m)}(cos\varphi)\times\nonumber\\
&&\times\int_{V}(r')^{n}\rho(r',\varphi',\lambda')P_{n}^{(m)}(cos\varphi')cosm(\lambda-\lambda')dv\nonumber
\end{eqnarray}
or
\begin{eqnarray}
U(r,\varphi,\lambda)=G\sum_{n=0}^{\infty}\frac{Y_{n}(\varphi,\lambda)}{r^{n+1}}\nonumber,
\end{eqnarray}
where
\begin{equation}\label{3}
Y_{n}(\varphi,\lambda)=\sum_{m=0}^{n}P_{n}^{(m)}(cos\varphi)[A_{nm}cosm\lambda+B_{nm}sinm\lambda].
\end{equation}
Here, $\lambda$ is longitude and $\varphi$ is the geocentric
latitude. The symbol V indicates that the integration should be
extended to the whole volume of the Earth. The coefficients $A_{nm}$
and $B_{nm}$ from (\ref{3}) are
\begin{eqnarray}\label{4}
A_{nm}=\frac{2}{\delta_{m}}\frac{(n-m)!}{(n+m)!}\int_{V}(r')^{n}P_{n}^{(m)}(cos\varphi')cos(m\lambda')\rho(r',\varphi',\lambda')dv\nonumber,\\
\\B_{nm}=\frac{2}{\delta_{m}}\frac{(n-m)!}{(n+m)!}\int_{V}(r')^{n}P_{n}^{(m)}(cos\varphi')sin(m\lambda')\rho(r',\varphi',\lambda')dv\nonumber,
\end{eqnarray}
where \begin{eqnarray}\
       \delta_{m}=\left\{\begin{array}{rcl}
       2,\ m\geq 1\\
       1,\ m=0\nonumber,
       \end{array}\right.
       \end{eqnarray}
while $P_{n}$ and $P_{n}^{(m)}$ are respectively the conventional
zonal harmonics of $n^{th}$ degree and the associated function of
Legendre of $n^{th}$ degree and $m^{th}$ order.

We mention that the recommended geopotential form by U.A.I.
\cite{MUE} is

\begin{eqnarray}\label{6}
U(r,\varphi,\lambda)&=&\frac{GM}{r}[1-\sum_{n=1}^{\infty}(\frac{a_{e}}{r})^{n}J_{n}P_{n}(cos\varphi)+\nonumber\\
\nonumber\\&&+\sum_{n=1}^{\infty}\sum_{m=1}^{\infty}(\frac{a_{e}}{r})^{n}P_{n}^{(m)}(cos\varphi)(C_{nm}cosm\lambda+S_{nm}sinm\lambda)]\nonumber,
\end{eqnarray}
where the harmonics coefficients of the geopotential are
\begin{eqnarray}\label{7}
J_{n}=\frac{1}{M{a_{e}}^{n}}\frac{2(n-m)!}{(n+m)!}\int_{V}(r')^{n}P_{n}^{(m)}(cos\varphi)cos(m\lambda')\rho(r',\varphi',\lambda')dv\nonumber,\\
C_{nm}=\frac{1}{M{a_{e}}^{n}}\frac{2(n-m)!}{(n+m)!}\int_{V}(r')^{n}P_{n}^{(m)}(cos\varphi')cos(m\lambda')\rho(r',\varphi',\lambda')dv,\\
S_{nm}=\frac{1}{M{a_{e}}^{n}}\frac{2(n-m)!}{(n+m)!}\int_{V}(r')^{n}P_{n}^{(m)}(cos\varphi')sin(m\lambda')\rho(r',\varphi',\lambda')dv\nonumber,
\end{eqnarray}
while
\begin{eqnarray}
P_{n}^{(m)}(cos\varphi)=\frac{1}{\sqrt{h_{nm}}}\sqrt{\frac{\delta_{n}(n-m)!}{(n+m)!}}\frac{d^{m}P_{n}(cos\varphi)}{d(cos\varphi)^{m}}sin^{m}\lambda\nonumber.
 \end{eqnarray}

There are three kinds of Legendrians in use: the conventional
Legendrian when $h_{nm}=\frac{\delta_{n}(n-m)!}{(n+m)!}$, the
normalized Legendrian when $h_{nm}=1$ and the fully normalized
Legendrian when $h_{nm}=(2n+1)^{-1}$ \cite{MUE}. We use the
conventional Legendrian.

The connection between the harmonics coefficients of the
geopotential from (\ref{7}) and the coefficients (\ref{4}) is given
by the relations\\
\begin{eqnarray}\label{9}
J_{n}&=&-\frac{1}{M{a_{e}}^{n}}A_{n0}\nonumber,\\
C_{nm}&=&\frac{1}{M{a_{e}}^{n}}A_{nm},\\
S_{nm}&=&\frac{1}{M{a_{e}}^{n}}B_{nm}\nonumber,
\end{eqnarray}
where $M$ is mass of the Earth.

Often, the geopotential is defined by the next expression:
\begin{eqnarray}
U(r,\varphi,\lambda)&=&\frac{GM}{r}[1+\sum_{n=1}^{\infty}(\frac{a_{e}}{r})^{n}I_{n}P_{n}(cos\varphi)+\nonumber\\
\nonumber\\&&+\sum_{n=1}^{\infty}\sum_{m=1}^{\infty}(\frac{a_{e}}{r})^{n}I_{nm}P_{nm}(cos\varphi)cosm(\lambda-\lambda_{nm})]\nonumber,
\end{eqnarray}
where the relations between the coefficients $I_{m}$, $I_{nm}$ and
the constants $\lambda_{nm}$ with the coefficients (\ref{7}) are
(see \cite{EK})
\begin{eqnarray}\label{11}
I_{n}&=&-J_{n}\nonumber,\\
I_{nm}&=&\sqrt{C_{nm}^{2}+S_{nm}^{2}}\nonumber,\\
\lambda_{nm}&=&\frac{1}{m}\arctan\frac{S_{nm}}{C_{nm}}\nonumber.
\end{eqnarray}

If instead the $P_{n}^{(m)}$ Legendre polynomials we consider the
functions $\overline{P}_{nm}$ , with
\begin{eqnarray}
\overline{P}_{nm}(x)=\sqrt{\frac{2(n-m)!(2n+1)}{(n+m)!}}P_{n}^{(m)}(x)\nonumber,
\end{eqnarray}
then the series (\ref{6}) becomes (see \cite{MUE})
\begin{eqnarray}
U(r,\varphi,\lambda)&=&\frac{GM}{r}[1-\sum_{n=1}^{\infty}(\frac{a_{e}}{r})^{n}J_{n}P_{n}(cos\varphi)+\nonumber\\
\nonumber\\&&+\sum_{n=1}^{\infty}\sum_{m=1}^{\infty}(\frac{a_{e}}{r})^{n}\overline{P}_{nm}(cos\varphi)(A_{nm}cosm\lambda+B_{nm}sinm\lambda)]\nonumber,
\end{eqnarray}
where
\begin{eqnarray}\
C_{nm}&=&A_{nm}\sqrt{\frac{2(n-m)!(2n+1)}{(n+m)!}}\nonumber,\\
\nonumber\\S_{nm}&=&B_{nm}\sqrt{\frac{2(n-m)!(2n+1)}{(n+m)!}}.\nonumber
\end{eqnarray}

If we note the following terms
\begin{eqnarray}\
q_{n0}&=&\sqrt{2n+1}\nonumber,\\
\nonumber\\q_{nm}&=&\sqrt{\frac{2(n-m)!(2n+1)}{(n+m)!}},\nonumber
\end{eqnarray}
then the polynomials $\overline{P}_{m}$ and $\overline{P}_{nm}$ may
be written as follows
\begin{eqnarray}\
\overline{P}_{n}(x)&=&-q_{n0}P_{n}(x)\nonumber,\\
\nonumber\\\overline{P}_{nm}(x)&=&q_{nm}P_{n}^{(m)}(x)\nonumber
\end{eqnarray}
and the series (\ref{6}) becomes (see \cite{MUE})
\begin{eqnarray}
U(r,\varphi,\lambda)&=&\frac{GM}{r}[1-\nonumber\\
\nonumber\\&&-\sum_{n=1}^{\infty}\sum_{m=0}^{\infty}(\frac{a_{e}}{r})^{n}\overline{P}_{nm}(cos\varphi)
(\overline{C}_{nm}cosm\lambda+\overline{S}_{nm}sinm\lambda)],\nonumber
\end{eqnarray}
where
\begin{eqnarray}
q_{n0}C_{n0}&=&I_{n}\nonumber,\\
q_{nm}\overline{C}_{nm}&=&C_{nm}\nonumber,\\
q_{nm}\overline{S}_{nm}&=&{S}_{nm}\nonumber.
\end{eqnarray}

The geopotential models are characterized by some constant values
(see Table 1), called the universal constants of geopotential. They
include the equatorial radius of the Earth ($a_{e}$), the geocentric
gravitational constant ($GM$) and the geometrical flattening of the
Earth ($f_{e}$).

In the following, it is necessary to know the dynamical flattening
of the Earth ($H$). We use the value of $H$ calculated from the
constant of precession \cite{MIH}
\begin{equation}
H=0.00327369\pm9.8\cdot10^{-8}.
\end{equation}

\begin{center}\begin{tabular}{|c|c|c|c|}
  \hline
  MODEL & $a_{e}[m]$ & $f_{e}$ & $GM\cdot10^{-14}[m^{3}\cdot s^{-2}]$ \\
  \hline
  SE-2 & 6378155.0 & $1/298.255$ & 3.986013 \\
  GEM-5, GEM-6 & 6378155 & $1/298.255$ & 3.986013 \\
  GEM-7, GEM-8 & 6378137.8 & $1/298.7925$ & 3.9860013 \\
  GEM-9, GEM-10 & 6378139.1 & $1/299.7925$ & $3.9860064\pm0.02$ \\
  EGM96  & 6378136.3 & $1/298.257$ & 3.986004415 \\
  \hline
\end{tabular}
\end{center}
\[  {\rm Table\ 1.} \ Universal\ constants\ of\ geopotential\]

\section{Determination of the moments of inertia of the Earth}

Using the expressions (\ref{4}) of the harmonics coefficients of the
geopotential, one finds the relations between the first coefficients
and the moments of inertia of the Earth $A'$, $B'$, $C'$, $D'$,
$E'$, $F'$ in the system $O\xi\eta\zeta$, namely
\begin{eqnarray}\label{20}
A_{20}&=&\frac{A'+B'}{2}-C'\nonumber,\\
A_{21}&=&E'\nonumber,\\
B_{21}&=&D',
\\A_{22}&=&\frac{B'-A'}{4} \nonumber,\\
B_{22}&=&\frac{F'}{2}.\nonumber
\end{eqnarray}

For the moments of inertia of the Earth, the following notations
were used:
\begin{eqnarray}\label{21}
A'&=&\int_{V}\rho(\xi,\eta,\zeta)(\eta^{2}+\zeta^{2})dv\nonumber,\\
B'&=&\int_{V}\rho(\xi,\eta,\zeta)(\xi^{2}+\zeta^{2})dv,\\
C'&=&\int_{V}\rho(\xi,\eta,\zeta)(\xi^{2}+\eta^{2})dv\nonumber,
\\D'&=&\int_{V}\rho(\xi,\eta,\zeta)\zeta\eta dv,\
E'=\int_{V}\rho(\xi,\eta,\zeta)\zeta\xi dv,\
F'=\int_{V}\rho(\xi,\eta,\zeta)\xi\eta dv,\nonumber
\end{eqnarray}
where $\rho$ is the density. On the other hand, from (\ref{9}) and
(\ref{20}) the following relations between the coefficients $I_{n}$,
$C_{nm}$, $S_{nm}$ and the moments of inertia of the Earth are
obtained:
\begin{eqnarray}\label{22}
Ma_{e}^{2}I_{2}&=&C'-\frac{A'+B'}{2}\nonumber,\\
4Ma_{e}^{2}C_{22}&=&B'-A'\nonumber,\\
 Ma_{e}^{2}C_{21}&=&E',
\\Ma_{e}^{2}S_{21}&=&D'\nonumber,\\
2Ma_{e}^{2}S_{22}&=&F'.\nonumber
\end{eqnarray}
The five relations (\ref{22}) are insufficient to determine the six
moments of inertia (\ref{21}) of the Earth.The system would be
complete if another independent equation is added.

Erzhanov and Kalybaev \cite{EK} had the idea to use the following
expression for the sixth equation of the system
\begin{eqnarray}
H'=\frac{1}{2C'}[2C'-(A'+B')]\nonumber,
\end{eqnarray}
taking the $H'=H$ without a motivation for this approximation. To
avoid it, Prof. I. Mihaila deducted an equation for calculating the
polar moment $C'$, using for this purpose the expression of $H$
\cite{MIH}.

With this equation of the polar moment:
\begin{equation}\label{24}
ax^{3}+bx^{2}+cx+d=0,
 \end{equation}
 where the coefficients $a$, $b$, $c$ and $d$ are respectively
\begin{eqnarray}\label{25}
a&=&8H^{3}\nonumber,\\
b&=&8H^{3}a'\nonumber,\\
c&=&-2H(3-4H)a'^{2}+2H(3-2H)^{2}(\frac{a'^{2}-b'^{2}}{4}-D'^{2}-E'^{2}-F'^{2})\nonumber,
\\d&=&-2(1-H)a'^{3}+(3-2H)^{2}a'(\frac{a'^{2}-b'^{2}}{4}-D'^{2}-E'^{2}-F'^{2})+
\nonumber
\\&&+(3-2H)^{3}(2D'E'F'+\frac{a'+b'}{2}E'^{2}+\frac{a'-b'}{2}D'^{2})\nonumber,\\
a'&=&-2J_{2}Ma_{e}^{2}\nonumber ,\\
b'&=&B'-A'=4Ma_{e}^{2}C_{22},\nonumber
\end{eqnarray}
the system of six independent algebraic equations (\ref{22}) and
(\ref{24}) for the six searched moments of inertia $A'$, $B'$, $C'$,
$D'$, $E'$, $F'$ is obtained.

If the values of harmonic coefficients of order $n = 2$ for the
geopotential and the dynamical flattening of the  Earth are known,
then the normalized moments of inertia $\overline{A}'$,
$\overline{B}'$, $\overline{C}'$, $\overline{D}'$, $\overline{E}'$,
$\overline{F}'$ can be determined easily. We note here
$\overline{A}'=A'/Ma_{e}^{2}$, etc. The $SE-2$ model is completed by
the $SE-2'$ model, where the harmonics coefficients
$\overline{C}_{21}=-0.001196\cdot10^{-6}$ and
$\overline{S}_{21}=-0.003466\cdot10^{-6}$ were calculated by
Erzhanov and Kalybaev \cite{EK}.
\begin{center}\begin{tabular}{|c|c|c|c|c|c|}
  \hline
  MODEL & $\overline{C}_{20}\cdot10^{6}$ & $\overline{C}_{21}\cdot10^{6}$ & $\overline{S}_{21}\cdot10^{6}$ & $\overline{C}_{22}\cdot10^{6}$ & $\overline{S}_{22}\cdot10^{6}$ \\
  \hline
  $SE-2$ & -484.16596 & 0.00000 & 0.00000 & 2.41290 & -1.36410 \\
  $SE-2'$ & -484.16596 & -0.001196 & -0.003466 & 2.41290 & -1.36410  \\
  $GEM-5$ & -484.16620 & -0.00120 & -0.00870 & 2.42820 & -1.36020 \\
  $GEM-6$ & -484.16610 & -0.00090 & -0.00120 & 2.42510 & -1.38830\\
  $GEM-7$ & -484.16460 & -0.00310 & -0.00090 & 2.43030 & -1.39460\\
  $GEM-8$ & -484.16460 & -0.00010 & -0.00030 & 2.43450 & -1.39530\\
  $GEM-9$ & -484.16555 & -0.00021 & -0.00406 & 2.43400 & -1.39786 \\
  $GEM-10$ & -484.16544 & -0.00104 & -0.00243 & 2.43404 & -1.39907 \\
  $EGM96$ & -484.16537 & -0.000187 & 0.001195 & 2.43914 & -1.40017 \\
  \hline
\end{tabular}
\end{center}
\[  {\rm Table\ 2.} \ Normalized\ harmonics\ coefficients\ of\ the\
geopotential\ (see \cite{EK}, \cite{LEM})\]
\begin{center}
\begin{tabular}{|c|c|c|c|}
  \hline
  MODEL & $\overline{A}'$ & $\overline{B}'$ & $\overline{C}'$ \\
  \hline
  SE-2 & 0.329619974 & 0.329626204 & 0.330705717 \\
  SE-2' & 0.329619974 & 0.329626204 & 0.330705717 \\
  GEM-5 & 0.329620259 & 0.329626529 & 0.330706023 \\
  GEM-6 & 0.329620507 & 0.329625671 & 0.330705717 \\
  GEM-7 & 0.329619038 & 0.329625314 & 0.330704801 \\
  GEM-8 & 0.329619033 & 0.329625319 & 0.330704801 \\
  GEM-9 & 0.329619643 & 0.329625927 & 0.330705412 \\
  GEM-10 & 0.329619643 & 0.329625927 & 0.330705412 \\
  EGM96 & 0.329619636 & 0.329625934 & 0.330705412 \\
  \hline
\end{tabular}
\end{center}
\[  {\rm Tables\ 3.a.} \ The\ normalized\ moments\ of\ inertia\ of\ the\
Earth\]
\begin{center}
\begin{tabular}{|c|c|c|c|}
  \hline
  MODEL & $\overline{D}'\cdot10^{6}$ & $\overline{E}'\cdot10^{6}$ & $\overline{F}'\cdot10^{6}$ \\
  \hline
  SE-2 & 0& 0 & -1.761045528 \\
  SE-2' & -0.004474587 & -0.001544029 & -1.761045528 \\
  GEM-5 & -0.011231652 & -0.001549193 & -1.756010649 \\
  GEM-6 & -0.001549193 & -0.001161895 & -1.792287593 \\
  GEM-7 & -0.001161895 & -0.004002083 & -1.800420858 \\
  GEM-8 & -0.000387298 & -0.000129099 & -1.801324554 \\
  GEM-9 & -0.005241437 & -0.000271109 & -1.804629500 \\
  GEM-10 & -0.003137117 & -0.001342634 & -1.806191603 \\
  EGM96 & 0.001543100 & -0.000241400 & -1.807607613 \\
  \hline
\end{tabular}
\end{center}
\[  {\rm Tables\ 3.b.} \ The\ normalized\ moments\ of\ inertia\ of\ the\
Earth\]
In Tables 3, using the models of geopotential from Table 2,
we evaluate these normalized moments.

Further, if the mass $M$ and the equatorial radius of the Earth are
known, then one can determine the central moments of inertia $A'$,
$B'$, $C'$, $D'$, $E'$, $F'$. In Table 4.a. and Table 4.b., using
the data from Table 1 and the value of the gravitational constant
given by the IAU (1976) System of Astronomical Constants, namely
$G=6.672\cdot10^{-11} m^{3}kg^{-1}s^{-2}$, these moments are
evaluated.

Once the values of the moments (\ref{21}) known, the principal
moments of inertia $A$, $B$, $C$ can be determined  by solving
 the secular equation (see \cite{EFIM}, \cite{VAL})

\begin{equation}\label{26}\Delta(q)=\left|
\begin{array}{clcr}
A'-q & -F' & -E'\\
-F' & B'-q & -D'\\
-E' & -D' & C'-q
\end{array}
\right|=0\end{equation}\\
 The roots $q_{1}$, $q_{2}$, $q_{3}$ of equation (\ref{26})
 represent the principal moments of inertia $A$, $B$, respectively
 $C$.\\
\begin{center}
 \begin{tabular}{|c|c|c|c|}
  \hline
  MODEL & $A'\cdot10^{-37} [kg\cdot m^{2}]$ & $B'\cdot10^{-37} [kg\cdot m^{2}]$ & $C'\cdot10^{-37} [kg\cdot m^{2}]$ \\
  \hline
  SE-2 & 8.010992630 & 8.011144042 & 8.037380227 \\
  SE-2' & 8.010992630 & 8.011144042 & 8.037380227 \\
  GEM-5 & 8.010999557 & 8.011151941 & 8.037387664 \\
  GEM-6 & 8.011005584 & 8.011131088 & 8.037380227 \\
  GEM-7 & 8.010903161 & 8.011055690 & 8.037291025 \\
  GEM-8 & 8.010903040 & 8.011055812 & 8.037291025 \\
  GEM-9 & 8.010931380 & 8.011084104 & 8.037319434 \\
  GEM-10 & 8.010931380 & 8.011084104 & 8.037319434 \\
  EGM96 & 8.010920187 & 8.011073251 & 8.037308375 \\
  \hline
\end{tabular}
\[  {\rm Tables\ 4.a.} \ The\ moments\ of\ inertia\ of\ the\
Earth\]
\end{center}
\begin{center}
\begin{tabular}{|c|c|c|c|}
  \hline
  MODEL & $D'\cdot10^{-30} [kg\cdot m^{2}]$ & $E'\cdot10^{-29} [kg\cdot m^{2}]$ & $F'\cdot10^{-32} [kg\cdot m^{2}]$ \\
  \hline
  SE-2 & 0& 0 & -4.279996317 \\
  SE-2' & -1.087491183 & -3.752566228 & -4.279996317\\
  GEM-5 & -2.729709620 & -3.765116622 & -4.267759686 \\
  GEM-6 & -0.376511662 & -2.823837454 & -4.355926167 \\
  GEM-7 & -0.282381394& -9.726470762 & -4.375656585 \\
  GEM-8 & -0.094127316 & -0.313757105 & -4.377852885\\
  GEM-9 & -1.273855995 & -0.658891032 & -4.385892467 \\
  GEM-10 & -0.762431051 & -3.263079405& -4.389688933 \\
  EGM96 & -0.375027747 & -0.586687176 & -4.393124297 \\
  \hline
\end{tabular}
\end{center}
\[  {\rm Table\ 4.b.} \ The\ moments\ of\ inertia\ of\ the\
Earth\]

\begin{center}
\begin{tabular}{|c|c|c|c|}
  \hline
  MODEL & $\overline{A}$ & $\overline{B}$ & $\overline{C}$ \\
  \hline
  SE-2 & 0.329619513 & 0.329626665 & 0.330705717 \\
  SE-2' & 0.329619513 & 0.329626665 & 0.330705717 \\
  GEM-5 & 0.329619801 & 0.329626987 & 0.330706023 \\
  GEM-6 & 0.329619939 & 0.329626239 & 0.330705717 \\
  GEM-7 & 0.329618555 & 0.329625797 & 0.330704801 \\
  GEM-8 & 0.329618549 & 0.329625803 & 0.330704801 \\
  GEM-9 & 0.329619167 & 0.329626403 & 0.330705412 \\
  GEM-10 & 0.329619160 & 0.329626410 & 0.330705412 \\
  EGM96 & 0.329619148 & 0.329626422 & 0.330705412 \\
  \hline
\end{tabular}
\end{center}
\[  {\rm Table\ 5.} \ The\ normalized\ principal\ moments\ of\ inertia\ of\ the\
Earth\]

The values of the Earth's normalized principal moments of inertia
$\overline{A}$, $\overline{B}$, $\overline{C}$, where
$\overline{A}=\frac{A}{Ma_{e}^{2}}$ etc., obtained by the solving of
the  equation (\ref{26}) with the geopotential models from Tables 3
are presented in Table 5. In Table 6, using the data from Table 5
and Table 1, are evaluated the principal moments of inertia $A$,
$B$, $C$.
\begin{center}
\begin{tabular}{|c|c|c|c|}
  \hline
  MODEL & $A\cdot10^{-37} [kg\cdot m^{2}]$ & $B\cdot10^{-37} [kg\cdot m^{2}]$ & $C\cdot10^{-37} [kg\cdot m^{2}]$ \\
  \hline
  SE-2 & 8.010981426 & 8.011155246 & 8.037380227 \\
  SE-2' & 8.010981426 & 8.011155246 & 8.037380227 \\
  GEM-5 & 8.010988425 & 8.011163072 & 8.037387664 \\
  GEM-6 & 8.010991779 & 8.011144893 & 8.037380227 \\
  GEM-7 & 8.010891423 & 8.011067429 & 8.037291025 \\
  GEM-8 & 8.010891277 & 8.011067575 & 8.037291025 \\
  GEM-9 & 8.010919812 & 8.011095672 & 8.037319434 \\
  GEM-10 & 8.010919642 & 8.011095842 & 8.037319434 \\
  EGM96 & 8.010908325 & 8.011085109 & 8.037308375 \\
  \hline
\end{tabular}
\end{center}
\[  {\rm Table\ 6.} \ The\ principal\ moments\ of\ inertia\ of\ the\
Earth\]

As seen from Tables 3 and 5 or from Tables 4 and 6, it is noticed
that $C'$ coincides with $C$. For all the nine geopotential models
used, the value of $C'$ found here coincides with the value of $C'$
obtained by considering \[H'=\frac{1}{2C'}[2C'-(A'+B')]=H\]  as in
the work \cite{EK}. It is thus demonstrated that the choice made by
Erzhanov and Kalybaev, namely $H'=H$, is valid until $10^{-9}$.

\section{Orientation of the ellipsoid of inertia}
Let $Oxyz$ be the system of the Earth's principal axes of inertia,
whose coordinate axes are chosen so that\\
\begin{eqnarray}\label{27}
A&=&\int_{V}\rho(x,y,z)(y^{2}+z^{2})dv\nonumber,\\
B&=&\int_{V}\rho(x,y,z)(z^{2}+x^{2})dv\nonumber,\\
C&=&\int_{V}\rho(x,y,z)(x^{2}+y^{2})dv\nonumber,\\
\nonumber\int_{V}\rho(x,y,z)xydv&=&\int_{V}\rho(x,y,z)xzdv=\int_{V}\rho(x,y,z)yzdv=0.\nonumber
\end{eqnarray}
 The orientation of the system with respect to $O\xi\eta\zeta$ may be given by
 the Euler angles. We use the notations from \cite{EK}
\begin{equation}\label{28}
\beta=\widehat{(O\xi, ON)},  \alpha=\widehat{(ON,Ox)},
\gamma=\widehat{(O\zeta, Oz)},
\end{equation}
where $ON$ is the intersection between the plans $O\xi\eta$ and
$Oxy$, called the line of nodes.

Let $(p_{\xi},p_{\eta},p_{\zeta})$, $(q_{\xi},q_{\eta},q_{\zeta})$,
$(r_{\xi},r_{\eta},r_{\zeta})$ be the direction cosines of the axes
$Ox$, $Oy$, $Oz$ in respect with $O\xi\eta\zeta$. They are the
projections of the unit vectors $\textbf{p}$, $\textbf{q}$,
$\textbf{r}$ of the principal axes in the system $O\xi\eta\zeta$. On
the other hand, the $O\xi\eta\zeta$ system overlaps $Oxyz$ by three
rotations $R_{\beta}$, $R_{\gamma}$, $R_{\alpha}$. The direction
cosine have the following expressions:
\begin{eqnarray}\label{30}
p_{\xi} &=&cos(x,\xi)= cos\beta cos\alpha  - sin\beta sin\alpha cos\gamma,\nonumber\\
p_{\eta}&=&cos(x,\eta)= sin\beta cos\alpha + cos\beta sin\alpha cos\gamma,\nonumber\\
p_{\zeta}&=&cos(x,\zeta)= sin\alpha sin\gamma,\nonumber\\
q_{\xi}&=&cos(y,\xi)= -cos\beta sin\alpha - sin\beta cos\alpha
cos\gamma,\nonumber\\
q_{\eta}&=&cos(y,\eta)=-sin\beta sin\alpha+cos\beta cos\alpha
cos\gamma,
\\q_{\zeta}&=&cos(y,\zeta)=cos\alpha sin\gamma,\nonumber\\
r_{\xi}&=&cos(z,\xi)=sin\beta sin\gamma,\nonumber\\
r_{\eta}&=&cos(z,\eta)=-cos\beta sin\gamma,\nonumber\\
r_{\zeta}&=&cos(z,\zeta)=cos\gamma\nonumber
\end{eqnarray}
and satisfy the orthogonality conditions
\begin{eqnarray}\label{31}
\textbf{p}\cdot\textbf{p}=\textbf{q}\cdot\textbf{q}=\textbf{r}\cdot\textbf{r}=1,
\\\textbf{p}\cdot\textbf{q}=\textbf{p}\cdot\textbf{r}=\textbf{q}\cdot\textbf{r}=0.\nonumber
\end{eqnarray}
The direction cosines are given by the relations
\begin{equation}\label{32}
\frac{\gamma_{i1}}{\delta_{i1}}=\frac{\gamma_{i2}}{\delta_{i2}}=\frac{\gamma_{i3}}{\delta_{i3}}=\frac{1}{\sqrt{\delta_{i1}^{2}+\delta_{i2}^{2}+\delta_{i3}^{2}}},
\end{equation}
where $\delta_{i1}, \delta_{i2}, \delta_{i3}$, with i=1,2,3, are the
cofactors of the elements in row $i$ of the determinant $\Delta$
which appears in the secular equation (\ref{26}), $q$ being
successively replaced by the principal moments of inertia $A$, $B$
and respectively $C$ (see \cite{EFIM}, \cite{MI}). In the relation
(\ref{32}), we have $(\gamma_{11},\gamma_{12}, \gamma_{13}
)=(p_{\xi},p_{\eta}, p_{\zeta})$, $(\gamma_{21},\gamma_{22},
\gamma_{23})=(q_{\xi},q_{\eta}, q_{\zeta})$,
$(\gamma_{31},\gamma_{32}, \gamma_{33})=(r_{\xi},r_{\eta},
r_{\zeta})$.

In determining the orientation of the principal axes of inertia,
besides the orthogonality conditions (\ref{31}), it is also required
to fulfill the condition that the system $Oxyz$ to have the same
orientation as the system $O\xi\eta\zeta$, namely (see \cite{EFIM})
\begin{equation}\label{33}
\left|
\begin{array}{clcr}
p_{\xi}& p_{\eta} & p_{\zeta}\\
q_{\xi} & q_{\eta} &  q_{\zeta}\\
r_{\xi} & r_{\eta} & r_{\zeta}
\end{array}
\right|=1.\end{equation}\\
Once determined the direction cosines, it can be obtained the three
Euler's angles by the relations (\ref{30}). Thus, for determining
the angle $\gamma$ it is used the following relation
\begin{eqnarray}
cos\gamma=r_{\zeta}\nonumber,
\end{eqnarray}
and to obtain the angle $\alpha$, the following relations are used
\begin{eqnarray}\label{35}
 p_{\zeta}=sin\alpha sin\gamma, \nonumber
\nonumber\\q_{\zeta}=cos\alpha sin\gamma  .\nonumber
\end{eqnarray}
For the angle $\beta$ we have the following formulae
\begin{eqnarray}\label{36}
 sin\beta sin\gamma=p_{\eta}q_{\zeta}-q_{\eta}p_{\zeta},\nonumber
\nonumber\\cos\beta sin\gamma=p_{\xi}q_{\zeta}-q_{\xi}p_{\zeta}
.\nonumber
\end{eqnarray}

If $\gamma=0$, then the axes $Oz$ and $O\zeta$ coincide  and the
plan $Oxy$ coincides with $O\xi\eta$. The problem of the orientation
for the system $Oxyz$ is reduced in this case  to the problem of the
orientation for the plan system $Oxy$ in relation to $O\xi\eta$. We
have from (\ref{31}):
\begin{eqnarray}\label{37}
p_{\xi}&=&cos\beta cos\alpha-sin\beta sin\alpha=cos(\beta+\alpha)=cos\lambda,\nonumber\\
p_{\eta}&=&sin\beta cos\alpha+cos\beta
sin\alpha=sin(\beta+\alpha)=sin\lambda,
\\q_{\xi}&=&-cos\beta sin\alpha-sin\beta cos\alpha=-sin(\beta+\alpha)=
-sin\lambda, \nonumber\\
q_{\eta}&=&-sin\beta sin\alpha+cos\beta
cos\alpha=cos(\beta+\alpha)=cos\lambda,\nonumber
\end{eqnarray}
where it was noted by $\lambda$ the angle between axes $Ox$ and
$O\xi$. The expressions (\ref{37}) give us the ellipse of inertia
orientation in the plan $O\xi\eta$.

The orientation of the ellipsoid of inertia corresponding to
geopotential models is given in Table 7.a and Table 7.b. From the
model $SE-2$ with $\gamma=0$ was determined the orientation of $Oxy$
in relation to $O\xi\eta$. This orientation is used as a standard
for the other models.

In Table 7.b, the longitudes and the geocentric latitudes of the
ellipsoid of inertia axes are also given. For the determination of
the coordinates $\lambda_{A}$ and $\varphi_{A}$ of $Ox$, we have the
following relations
\begin{eqnarray}
 p_{\xi}&=&cos\lambda_{A} cos\varphi_{A},\nonumber
\nonumber\\p_{\eta}&=&sin\lambda_{A} cos\varphi_{A},\nonumber\\
p_{\zeta}&=&sin\varphi_{A}.\nonumber\
\end{eqnarray}
Similarly, for the determination of the coordinates $\lambda_{B}$
and $\varphi_{B}$ of $Oy$, we have the following relations
\begin{eqnarray}\label{39}
 q_{\xi}&=&cos\lambda_{B} cos\varphi_{B},\nonumber
\nonumber\\q_{\eta}&=&sin\lambda_{B} cos\varphi_{B},\nonumber\\
q_{\zeta}&=&sin\varphi_{B}.\nonumber
\end{eqnarray}

For $Oz$, the longitude $\lambda_{C}$ is determined from the
following formula
\begin{eqnarray}
\lambda_{C}=360^{\circ}-(90^{\circ}+\beta)\nonumber,
\end{eqnarray}
where $\beta$ is given by (\ref{28}).
\begin{center}
 \begin{tabular}{|c|c|c|c|}
  \hline
   MODEL & $\alpha^{\circ}$ & $\beta^{\circ}$ & $\gamma$ (in arcsec ("))  \\
  \hline
  SE-2 & - & - & 0 \\
  SE-2'& 4.3 & -19 & 0.9 \\
  GEM-5 & -7 & -7.9 & 2.2 \\
  GEM-6 & 19.4 & -36.8 & 0.4 \\
  GEM-7 & 58.7 & -73.6 & 0.8 \\
  GEM-8 & 3.5 & -18.4 & 0.1 \\
  GEM-9 & -11.9 & -3.0 & 1.0 \\
  GEM-10 & 8.2 & -23.1 & 0.7 \\
  EGM96 & 23.7 & -38.6 & 0.3 \\
  \hline
\end{tabular}
\end{center}
\[{\rm Table\ 7.a.} \ The\ orientation\ of\ the\
ellipsoid\ of\ inertia\]

\begin{center}
\begin{tabular}{|c|c|c|c|c|c|c|}
  \hline
  MODEL & ${\lambda_{A}}^{\circ}$ & ${\varphi_{A}}^{\circ}$ & ${\lambda_{B}}^{\circ}$ & ${\varphi_{B}}^{\circ}$ & ${\lambda_{C}}^{\circ}$ & $(\frac{\pi}{2}-\varphi_{C})"$ \\
  \hline
  SE-2 & -14.7 & 0 & 75.3 & 0 & 0 & 0 \\
  SE-2' &-14.7 & $1.9\cdot10^{-5}$ & 75.3 & $2.5\cdot10^{-4}$ & 251 & 0.9 \\
  GEM-5 & -14.9 & $-7.3\cdot10^{-5}$ &75.1 & $5.97\cdot10^{-4}$ & 262.1 & 2.2 \\
  GEM-6 & -17.4 & $3.4\cdot10^{-5}$ & 72.6& $9.7\cdot10^{-5}$ & 233.2 & 0.4 \\
  GEM-7 & -14.9 & $1.9\cdot10^{-4}$ & 75.1 & $1.1\cdot10^{-4}$ & 196.4 & 0.8 \\
  GEM-8 & -14.9 & $1.3\cdot10^{-6}$ & 75.1 & $2.2\cdot10^{-5}$ & 251.6 & 0.1 \\
  GEM-9 & -14.9 & $-5.7\cdot10^{-5}$ & 75.1 & $2.7\cdot10^{-4}$ & 267.0 & 1.0 \\
  GEM-10 & -14.9 & $2.6\cdot10^{-5}$ & 75.1 & $1.8\cdot10^{-4}$ & 246.9 & 0.7 \\
  EGM96 & -14.9 & $3.3\cdot10^{-5}$ & 75.1& $7.6\cdot10^{-5}$ & 231.4 & 0.3 \\
  \hline
\end{tabular}

\end{center}
\[{\rm Table\ 7.b.} \ The\ orientation\ of\ the\
ellipsoid\ of\ inertia\]

As it is observed in Table 7.b, for the geopotential models
considered, the longitude of $Ox$ axis of the triaxial ellipsoid of
inertia is about $-15^{\circ}$ and the longitude of  $Oy$ axis has a
value close to $75^{\circ}$, except for the model $GEM-6$ for which
$\lambda_{A}\simeq-17^{\circ}$ and $\lambda_{B}\simeq73^{\circ}$.
From the values obtained, except for the model GEM-5, it is
determined the mean ellipsoid of inertia with the principal moments
$A= 8.010935639\cdot10^{37}kg\cdot m^{2}$, $B=
8.011108377\cdot10^{37}kg\cdot m^{2}$, $C=
8.037333747\cdot10^{37}kg\cdot m^{2}$ and with the orientation
$\lambda_{A}=-15^{\circ}.2$, $\lambda_{B}=74^{\circ}.8$. The value
of $H$ obtained from the mean values of the principal moments of
inertia, namely $H=0.00327369$, is the same as in \cite{MIH}.

In Figure 1, the position of the inertial pole on the surface of the
Earth is represented with respect to the conventional international
pole $P_{0}$ for the nine models of geopotential considered. Since
the angle $\gamma$ is a small angle, then, in a Cartesian reference
$XP_{0}Y$ in the tangent plan at $P_{0}$, with the axis ${P_{0}X}$
tangent to the Greenwich meridian, the pole  of inertia $P_{i}$ will
have the polar coordinates ($\lambda_{C}$, $\gamma$).

It is remarked that the pole of the ellipsoid SE-2 when $\gamma=0$
coincides with the conventional international pole $P_{0}$, the
other  positions of the pole remaining in the neighborhood, except
for the pole of inertia corresponding to $GEM-5$. The coordinates of
the mean pole $\overline{P}$ are $\lambda_{C}=209^{\circ}.7$ and
$\gamma=0".5$. Therefore the mean polar axis differs from the
rotation axis by $0".5$. The mean pole $\overline{P}$ deviates
approximately by $15$ meters from the pole of rotation.

\scalebox{1}

\begin{pspicture}(0,-4.04)(8.854688,4.02)
\psline[linewidth=0.04cm,arrowsize=0.05291667cm
2.0,arrowlength=1.4,arrowinset=0.4]{->}(0.0,-0.52)(8.44,-0.54)
\psline[linewidth=0.04cm,arrowsize=0.05291667cm
2.0,arrowlength=1.4,arrowinset=0.4]{->}(5.2,4.0)(5.2,-3.78)
\usefont{T1}{ptm}{m}{n}
\rput(8.685625,-0.41){$Y$}
\usefont{T1}{ptm}{m}{n}
\rput(5.59625,-3.83){$X$, $G$} \psdots[dotsize=0.12](5.2,-0.53)
\psline[linewidth=0.04cm](3.22,-0.44)(3.22,-0.62)
\psline[linewidth=0.04cm](1.22,-0.44)(1.22,-0.6)
\psline[linewidth=0.04cm](4.2,-0.42)(4.2,-0.6)
\usefont{T1}{ptm}{m}{n}
\rput(4.2815623,-0.85){0",5}
\usefont{T1}{ptm}{m}{n}
\rput(3.2304688,-0.85){1"}
\usefont{T1}{ptm}{m}{n}
\rput(1.2276562,-0.81){2"}
\psline[linewidth=0.04cm](5.14,1.46)(5.28,1.46)
\usefont{T1}{ptm}{m}{n}
\rput(5.570469,1.49){1"} \psdots[dotsize=0.12](5.02,1.24)
\psdots[dotsize=0.12](5.02,-0.42) \psdots[dotsize=0.12](4.62,-0.18)
\psdots[dotsize=0.12](4.42,0.04) \psdots[dotsize=0.12](3.96,-0.02)
\psdots[dotsize=0.12](3.22,-0.28) \psdots[dotsize=0.12](3.42,-0.12)
\psdots[dotsize=0.12](1.2,-0.16)
\pscircle[linewidth=0.04,dimen=outer](4.25,0.54){0.1}
\usefont{T1}{ptm}{m}{n}
\rput(4.3907814,0.89){$\overline{P}$}
\end{pspicture}

\[{\rm Figure}\ 1.\ The\ position\ of\ the\ inertial\ pole\ on\ the\ surface\ of\
the\
Earth \]

The results obtained show that the approximation made in the paper
\cite{EK} is satisfied and the improved values for the principal
moments of inertia $A$, $B$, $C$ are obtained. On the other hand, it
is better emphasized the fact that the polar axis of inertia is
located in the neighborhood of the Earth's rotation axis. For the
geopotential models considered, the longitudes of the axes $Ox$ and
$Oy$ of the
triaxial ellipsoids of inertia have concordant values.\\
\textbf{Acknowledgements}. I wish to express my deep gratitude to my
advisor, Professor Ieronim Mihaila from the University of Bucharest,
who encouraged and assisted me in the development and completion of
this paper.

\emph{Alina-Daniela V\^{\i}lcu}\\
Petroleum-Gas University of Ploiesti, Department of Mathematics,\\
Bulevardul Bucuresti, Nr. 39, Ploiesti 100680, Romania\\
E-mail: viallin@yahoo.com

\end{document}